\documentclass[a4paper,10pt,aps,pre,twocolumn,superscriptaddress,amsmath,amssymb,nofootinbib]{revtex4}
\usepackage{graphicx,color,soul}
\usepackage[colorlinks=true,linkcolor=blue]{hyperref}%

\newcommand{\mean}[1]{\langle #1 \rangle}

\renewcommand{\vec}[1]{\mathbf #1}

\usepackage[dvipsnames]{xcolor}

\newcommand{\ep}{\epsilon}

\newcommand{\kap}{\kappa}

\newcommand{\sig}{\sigma}
\newcommand{\kT}{k_\text{B}T}

\begin{document}

\title{Is directed percolation in colloid-polymer mixtures linked to dynamic arrest?}

\author{David Richard}
\affiliation{Institut f\"ur Physik, Johannes Gutenberg-Universit\"at Mainz, Staudingerweg 7-9, 55128 Mainz, Germany}
\author{C. Patrick Royall}
\affiliation{HH Wills Physics Laboratory, Tyndall Avenue, Bristol BS8 1TL, UK}
\affiliation{School of Chemistry, University of Bristol, Cantock's Close, Bristol, UK}
\affiliation{Centre for Nanoscience and Quantum Information, Tyndall Avenue, Bristol, UK}
\author{Thomas Speck}
\affiliation{Institut f\"ur Physik, Johannes Gutenberg-Universit\"at Mainz, Staudingerweg 7-9, 55128 Mainz, Germany}

\begin{abstract}
  Using computer simulations, we study the dynamic arrest in a schematic model of colloid-polymer mixtures combining short-ranged attractions with long-ranged repulsions. The arrested gel is a dilute rigid network of colloidal particles bonded due to the strong attractions. Without repulsions, the gel forms at the spinodal through arrested phase separation. In the ergodic suspension at sufficiently high densities, colloidal clusters form temporary networks that percolate space. Recently [Nat. Commun. \textbf{7}, 11817 (2016)], it has been proposed that the transition of these networks to directed percolation coincides with the onset of the dynamic arrest, thus linking structure to dynamics. Here, we evaluate for various screening lengths the underlying gas-liquid binodal and the percolation transitions. We find that directed percolation shifts the continuous percolation line to larger densities, but even beyond this line the suspension remains ergodic. Only when approaching the spinodal does dynamic arrest occur. Competing repulsions thus do not modify the qualitative scenario for non-equilibrium gelation, although the structure of the emerging percolating network shows some differences.
\end{abstract}

\maketitle


After preparation, many soft materials do not reach their thermodynamically stable state but are dynamically arrested~\cite{pusey1987observation,weeks2000three,cipeletti2005,lu2008gelation,royall2008direct,eberle2011dynamical,ruzicka2011observation,hunter2012physics,pinchaipat2017experimental}. One example is low-density colloidal suspensions with short-range attractive forces, which form a gel, a non-equilibrium network structure of bonded particles~\cite{zaccarelli2007colloidal}. For colloid-polymer mixtures in which the polymers induce entropic depletion forces between the colloidal particles, there is now ample evidence that percolation~\cite{stauffer1994introduction} is necessary but not sufficient, and that gelation is related to liquid-gas phase separation that is arrested~\cite{verhaegh1997,poon2002,cates2004theory,manley2005glasslike,lu2008gelation}. This arrest is caused mainly by the large cost of breaking bonds and the high density of the colloidal liquid phase, although hydrodynamics also plays a role~\cite{royall2015probing,tsurusawa2018}. This scenario is supported by experiments directly imaging and tracking the colloidal particles through confocal microscopy~\cite{lu2008gelation,royall2008direct,royall2018vitri}, and corroborated by simulations of systems with short-ranged attractions~\cite{lu2008gelation,griffiths2017local,royall2018vitri,richard2018}. These systems are characterized by a metastable critical point terminating gas-liquid coexistence within the gas-solid two-phase region~\cite{anderson2002}.

For suspensions of nanocolloids with thermosensitive molecular brushes, an alternative scenario has been proposed in which the gelation line is located before the phase separation and at higher densities is linked to the location of the attractive glass~\cite{eberle2011dynamical,eberle2012dynamical}. In a numerical study~\cite{valadez2013dynamical} of the adhesive hard-sphere model~\cite{baxter1968percus} this gelation line has been related to the mean-field rigidity transition~\cite{he1985elastic}. For colloidal particles with additional electrostatic repulsions, the onset of directed percolation (DP) has been proposed as a structural transition taking place concurrently with gelation~\cite{kohl2016directed}. In contrast to continuous percolation, in the case of DP only forward paths along an arbitrary direction are considered~\cite{hinrichsen2000non}. Moreover, in computer simulations of sticky spheres it has been demonstrated that adding a screened electrostatic potential shifts the percolation line~\cite{valadez2013percolation}. Hence, while for short-ranged attractions the specific shape of the pair potential is known to be irrelevant, adding a competing long-range repulsive term might play a role in determining the location and microscopic mechanism of gelation. In this Communication, we study such a model potential. However, one should bear in mind that for the important class of experiments which use confocal microscopy to study colloidal systems in 3d real space (so-called particle-resolved studies) simple addition of spherically symmetric attractions and repulsions does not seem to hold~\cite{royall2018hunting}.

\begin{figure}[t!]
  \includegraphics[width=\linewidth]{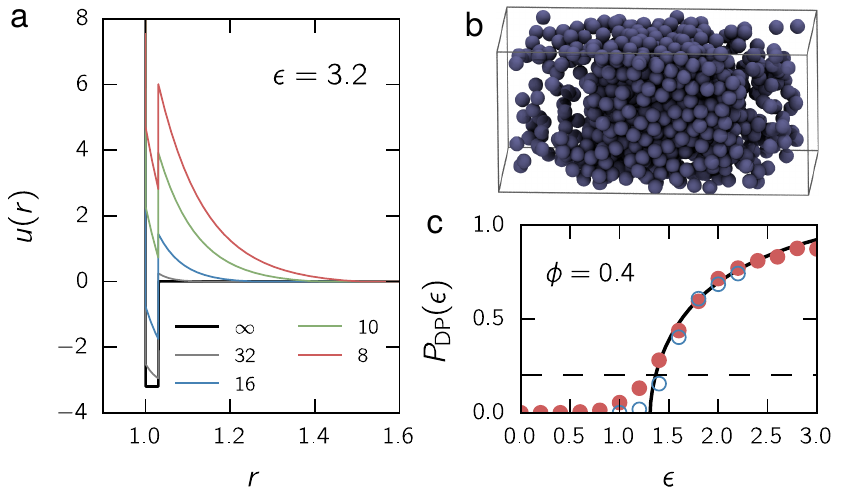}
  \caption{\textbf{Model and methods.} (a)~Pair potential for $\ep=3.2$ and various values for $\kappa$. (b)~Simulation snapshot in the slab geometry used to compute coexisting densities. (c)~Probability $P_{\text{DP}}$ that a particle participates in a directed path as a function of the attraction strength $\ep$, $\kappa\to\infty$, and $\phi=0.4$ for two system sizes: $N=1000$ (filled red symbols) and $N=10,000$ (empty blue symbols). The solid line is a fit to $P_{\text{DP}}\propto(\ep-\ep_{\text{DP}})^{\beta_{\text{DP}}}$ (\emph{cf.} main text). The dashed line indicates the criteria $P_{\text{DP}}>0.2$ for points included in the fit.}
  \label{fig:model}
\end{figure}
\begin{figure}[h!]
  \includegraphics[width=\linewidth]{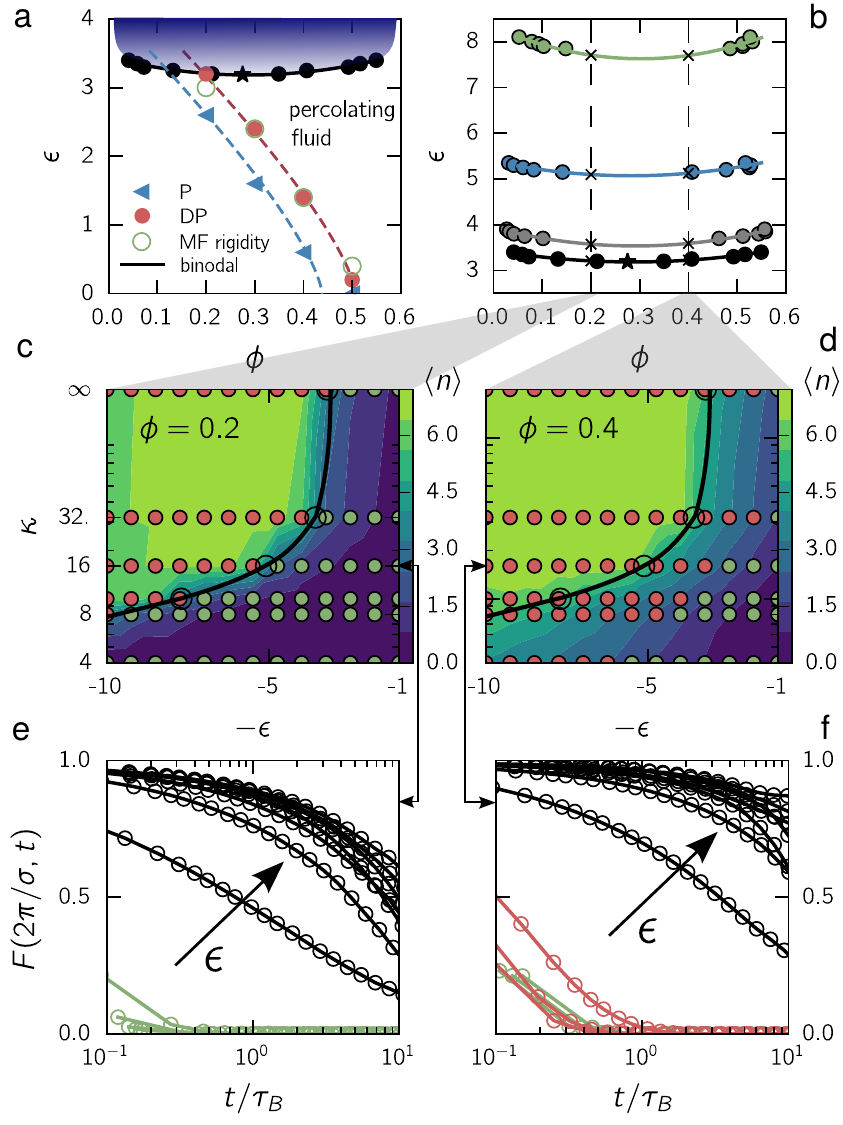}
  \caption{\textbf{Statics and dynamics.} (a)~Phase diagram of the SW fluid ($\kappa\to\infty$) in the plane ($\phi,\ep$). The black line shows the binodal with symbols showing measured coexisting densities and the black star indicating the critical point. Also shown are the threshold attractions $\ep_\text{P}$ for continuous percolation (P, blue triangles) and $\ep_\text{DP}$ for directed percolation (DP, red discs) to occur. The mean-field rigidity transition $\mean{n}=2.4$ (empty green symbols) coincides with DP. The dashed red line is a guide to the eye and crosses the binodal at $\phi\simeq0.2$. (b)~Decreasing $\kap$ shifts the binodal to larger $\ep$. The different colors from black (bottom) to green (top) correspond to $\kappa=\infty,32,16,10$. (c,d)~Phase diagram in the plane ($-\ep,\kappa$) for two different packing fractions (c)~$\phi=0.2$ and (d)~$\phi=0.4$. The color map indicates the average number of bonds $\mean{n}$. Directed percolation is absent at green symbols ($P_{\text{DP}}<0.2$) and present at red symbols ($P_{\text{DP}}\geqslant0.2$). The black empty circles indicate the position of the binodal from the intersection in (b) of dashed lines and binodals (crosses). The black lines are guides to the eye. (e,f)~Self-intermediate scattering function $F(k,t)$ for $\kappa=16$ varying $\ep$ uniformly from $1$ to $10k_BT$ at two different packing fractions: (e)~$\phi=0.2$ and (f)~$\phi=0.4$. Color code: green if $P_{\text{DP}}<0.2$, red if $P_{\text{DP}}\geqslant0.2$, and black if the system has crossed the binodal.}
  \label{fig:results}
\end{figure}
\begin{figure}[t]
  \centering
  \includegraphics[width=\linewidth]{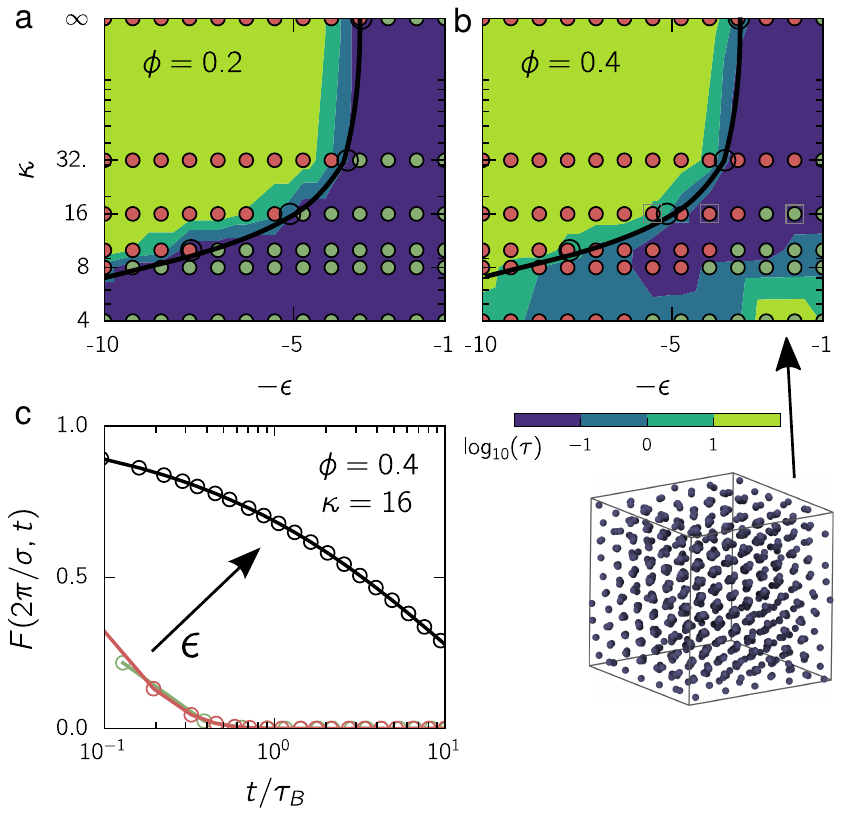}
  \caption{\textbf{Dynamic crossover.} Shown is the same plot as in Fig.~\ref{fig:results}(c,d) for (a)~$\phi=0.2$ and (b)~$\phi=0.4$ but now with the color indicating the structural relaxation time $\tau$. In (b), for large repulsions we observe the formation of a Wigner crystal (see snapshot). (c)~Same plot as Fig.~\ref{fig:results}(f) but now for a larger system with $N=10,000$ at three different $\ep$ indicated by gray squares in (b). Data for $N=1000$ is indistinguishable.}
  \label{fig:crossover}
\end{figure}

We study a system composed of $N$ particles, the diameters $\sig_i$ of which are drawn from a Gaussian distribution with mean $\sig$ corresponding to a polydispersity of $5\%$. Our pair potential reads $u(r)=u_\text{SW}(r)+u_\text{YK}(r)$, where the first contribution is the square well (SW) potential
\begin{equation}
  u_\text{SW}(r) =
  \begin{cases}
    \infty & \text{if $r \le \sigma_{ij}$}\\
    -\ep & \text{if $\sigma_{ij} < r <\sigma_{ij}+\delta$}\\
    0 & \text{if $r \ge \sigma_{ij}+\delta $}
  \end{cases}
\end{equation}
with $\sigma_{ij}=(\sigma_i+\sigma_j)/2$ modeling hard-core repulsion plus a short-range attraction. This part of the potential is fixed by two parameters: the attraction range $\delta$ and the attraction strength $\ep$. To be consistent with our previous experimental and numerical study~\cite{richard2018}, we set $\delta=0.03\sigma$. To model screened electrostatic interactions, we employ the Yukawa potential
\begin{equation}
  u_\text{YK}(r) = 
  C\left(\frac{2}{2+\kappa\sigma_{ij}}\right)^2\left(\frac{\sigma_{ij}}{r}\right)
  \exp[-\kappa(r-\sigma_{ij})]
\end{equation}
with screening length $\kappa^{-1}$ controlled in experiments by the salt concentration. The prefactor is set to $C=200\kT$ in agreement with Ref.~\cite{kohl2016directed}. In Fig.~\ref{fig:model}(a), we plot the total pair potential $u(r)$ for several values of $\kappa$ at $\ep=3.2\kT$, which is very close to the critical attraction strength of the SW fluid~\cite{richard2018}. For $\kap\to\infty$, we recover the SW fluid. In the opposite limit $\kappa\to0$ of unscreened charges, the system will form a Wigner crystal due to the (effective) large packing fraction~\cite{lindsay1982elastic} (cf. Fig.~\ref{fig:crossover}). Although more stylized, we shall see in Fig.~\ref{fig:results}(c) that this potential reproduces the same phase behavior as the model studied in Ref.~\cite{kohl2016directed}. To summarize, the model is characterized by three control parameters: the global packing fraction (colloid concentration), the attraction strength $\ep$ (related to polymer concentration), and the inverse screening length $\kappa$ (related to salt concentration).

We perform Monte Carlo simulations of $N$ particles (mostly $N=1000$) at fixed volume $V$ and temperature $T$. We employ only local moves with uniform displacements in the range $[-\delta l,\delta l]$ in each direction. We keep the acceptance probability for the local moves close to one half through adapting $\delta l$. The density is measured through the mean packing fraction $\phi=\pi\sig^3 N/(6V)$. We cut off the potential at $r_c=4/\kappa$ and we shift the Yukawa contribution by $u_\text{YK}(r_c)$ to enforce zero energy at the cutoff. In the following, we employ dimensionless lengths in units of $\sig$ and energies in units of $\kT$.

We perform two different types of simulations. We first study the equilibrium coexistence between a dilute gas and the dense liquid in a slab geometry with box lengths $L_x=L_y=L_z/2$, see Fig.~\ref{fig:model}(b). We first prepare a random hard-sphere configuration without overlaps in a cubic box two times smaller than the final volume at density $\phi=2\times\phi_c$, with $\phi_c=0.275$ the critical packing fraction of the SW model at $\delta=0.03$~\cite{largo2008vanishing,richard2018}. We then let the system equilibrate for $3\times10^7$ Monte Carlo sweeps and compute the density profile $\phi(z)$ along the $z$-axis for another $10^7$ sweeps. We fit the measured density profile to the mean-field expression
\begin{equation}
  \phi(z) = \frac{\phi_l-\phi_g}{2}
  + \frac{\phi_l-\phi_g}{2}\tanh\left(\frac{z-z_0}{2w}\right).
  \label{eq:profile}
\end{equation}
Here, $\phi_g$ and $\phi_l$ are the coexisting densities of the gas and liquid phase, respectively, and the interface position and width are $z_0$ and $w$. We perform four independent runs to calculate averaged density profiles.

The second type of simulations are performed in a cubic box of edge length $L$, which all start from a disordered initial configuration without overlaps. Here, we employ Kinetic Monte Carlo (KMC) simulations using the procedure described in Ref.~\cite{sanz2010dynamic}. Displacements in each direction are in the range [$-\delta l,\delta l$] with $\delta l=0.015\sigma$. We map the KMC dynamics onto Brownian Dynamics (with time step $\Delta t$, Brownian time $\tau_B=\sigma^2/D_0$, and $D_0$ the bare diffusion coefficient) through $\Delta t/\tau_B=p_a\delta l^2/(6\sigma^2)$ monitoring the acceptance probability $p_a$. We equilibrate the system for $2\times10^6$ sweeps and perform an analysis for an additional $3\times10^6$ sweeps. For ergodic suspensions, the typical relaxation time is of the order of $10^4$ Monte Carlo sweeps and corresponds to less than $0.1\tau_B$ after rescaling, which is consistent with Ref.~\cite{lopez2013dynamic} for hard-sphere suspensions below the freezing point. We construct a network of mutually bonded particles, whereby a bond between particles $i$ and $j$ is formed if their distance obeys $r_{ij}<\sigma_{ij}+\delta$ (\emph{i.e.}, they are within the range of the attractive well of the SW potential). We then compute three different quantities: the average number $\mean{n}$ of bonds formed with other particles, and the probabilities $P_\text{P}$ and $P_{\text{DP}}$ that a particle participates in a continuous and directed percolating path, respectively. For the latter, we follow closely the procedure described in Ref.~\cite{kohl2016directed}: We fix an arbitrary direction $\vec{d}$ and define a new bond network, where two particles $i$ and $j$ are now bonded if, additionally to our previous criteria, $\vec{d}\cdot\vec{r}_{ij}>0$ is obeyed. We then find for each particle whether it participates in a directed path of projected length $l_{\text{DP}}\ge L$. We calculate the probability $P_{\text{DP}}$ by averaging over all particles and configurations. One can extract the threshold $\ep_\text{DP}$ at which directed percolation sets in through fitting $P_{\text{DP}}(\ep)$ with the functional form $P_{\text{DP}}\propto(\ep-\ep_{\text{DP}})^{\beta_{\text{DP}}}$ with critical exponent $\beta_{\text{DP}}=0.58$~\cite{hinrichsen2000non}. In practice, to circumvent the smoothening of the transition caused by finite size effects, we only fit data points with $P_{\text{DP}}>0.2$ (cf. Ref.~\cite{kohl2016directed}). In Fig.~\ref{fig:model}(c), we show such a procedure for $\kappa\to\infty$ and $\phi=0.4$.


In Fig.~\ref{fig:results}(a), we plot the metastable gas-liquid binodal formed by the coexisting densities $\phi_{g,l}$ extracted from the Monte Carlo simulations for the SW model ($\kap\to\infty$) with the critical point at $\ep_c\simeq3.2$ and $\phi_c\simeq0.275$~\cite{largo2008vanishing}. Recently~\cite{richard2018}, we have confirmed experimentally and numerically (through a mapping onto the SW model) that gelation occurs along the spinodal, which for short-ranged attractive systems is often very close to the binodal~\cite{royall2018vitri}. We observe gelation for a wide range of densities $0.1<\phi<0.4$ with the lower density limit set by the onset of a percolating network of bonded particles [cf. Fig.~\ref{fig:results}(a)]. One should note that, while the SW model reproduces the onset of gelation, it does not show a true dynamic arrest but a crossover to a regime with slow (aging) dynamics. The line $\ep_{\text{DP}}(\phi)$ where directed percolation sets in has the same shape as for continuous percolation but is shifted to larger packing fractions. We find that the mean-field rigidity transition determined as the average number of bonds $\mean{n}=2.4$ agrees with the onset of directed percolation. As the global packing fraction increases, the DP threshold $\ep_{\text{DP}}$ decreases and goes to zero in the liquid-solid coexistence region of the hard-sphere fluid ($\ep=0$). In the opposite limit in the low density region, we find that the DP transition and the binodal intersect around $\phi\simeq0.2$. Hence, on one hand for packing fractions $0.1<\phi<0.2$ we have the formation of a gel without a structural signature. On the other hand, for $\phi>0.2$ we find that state points below the binodal but with $\ep>\ep_{\text{DP}}$ remain fully ergodic, \emph{i.e.}, form a percolating fluid. Only state points quenched through the binodal show dynamic arrest in agreement with previous work~\cite{toledano2009colloidal}. Hence, over a wide region of the phase diagram directed percolation is not associated with a pronounced change in the dynamics.

The central result of this Communication is that this picture remains essentially unchanged as we decrease $\kap$ thus increasing the range of the competing repulsions. In Fig.~\ref{fig:results}(b), we plot the phase diagram varying the inverse screening length $\kappa$. When increasing the repulsion strength between particles, one has to quench the system deeper (increasing $\ep$) to observe the metastable gas-liquid binodal. Note that the form of the binodal remains rather flat with a high-density liquid phase. In Fig.~\ref{fig:results}(c) and (d), we show a different cut ($-\ep,\kappa$) through parameter space now holding the packing fraction fixed. For $\phi=0.2$ shown in Fig.~\ref{fig:results}(c), we obtain a very similar phase diagram as reported in Ref.~\cite{kohl2016directed}. Interestingly, we find for all values of $\kappa$ that the DP transition coincides exactly with the position of the binodal. In contrast, when increasing the packing fraction to $\phi=0.4$ [Fig.~\ref{fig:results}(d)], we find no correlation between directed percolation and the location of the phase boundary. Note that also for finite $\kap$ we observe that directed percolation coincides with the rigidity transition $\mean{n}\simeq2.4$, hence, as seen for sticky spheres, the two transitions are intertwined. We remark that for strong repulsions the gas-liquid coexistence might terminate~\cite{archer2007}. However, for the values of $\kappa\geq 4$ considered here we do observe (meta)stable coexistence in our slab simulations.

To obtain insight into the dynamic behavior, we fix $\kappa=16$ (which would correspond to fixing the salt concentration) and progressively increase the attraction strength (increasing the polymer concentration). We record the self-intermediate scattering function (ISF)
\begin{equation}
  F(k,t) = \frac{1}{N}\sum_{i=1}^{N}
  \exp\left\{\mathrm i\vec{k}\cdot[\vec{r}_i(t)-\vec{r}_i(0)]\right\}
\end{equation}
at wave vector $k=2\pi/\sigma$. In Fig.~\ref{fig:results}(e,f) we plot the result for the two packing fractions. We observe a distinct jump of the shape of the ISF and an increase of the relaxation time $\tau$ (measured as $F(k,\tau)=1/e$) by about 2 orders of magnitude for $\phi=0.2$ and about one order of magnitude for $\phi=0.4$ between two successive value for $\ep$ exactly when crossing the binodal. This observation supports the scenario that dynamic arrest of the network coincides with the onset of phase separation. Specifically, for $\phi=0.4$ we show that the system remains ergodic (the intermediate structure function decays to zero) when crossing the directed percolation line (although the relaxation time does increase).

In Fig.~\ref{fig:crossover}, we compare the structural relaxation time $\tau$ with both the binodal and the DP transition. For both densities, down to $\kappa\simeq10$ we find that the binodal bounds the slow dynamics (which we identify with $\tau>10\tau_B$). For the lower density $\phi=0.2$, the region between slow dynamics and fast dynamics ($\tau<0.1\tau_B$) is narrow and broadens considerably for $\phi=0.4$. However, in this region the fluid remains ergodic and there is still a narrow band in which the relaxation time jumps by about one order of magnitude [Fig.~\ref{fig:crossover}(c) and Fig.~\ref{fig:results}(f)]. In contrast to $\phi=0.2$, state points characterized by directed percolation now extend far into the ergodic fluid. This indicates a highly ramified network in which bonds constantly reorganize, a percolating fluid.


In this Communication, we have reported simulations of a minimal model for colloid-polymer mixtures with competing short-range attractions and long-range repulsions. This model is characterized by three main parameters: the global packing fraction $\phi$ of colloidal particles, the attraction strength $\ep$, and the screening length $\kap^{-1}$. We found that the mean-field rigidity transition and directed percolation occur at the same location for various packing fractions, attraction strengths, and also screening lengths. Monitoring the dynamics through the self-intermediate scattering function, we have demonstrated that gelation in this model system is still controlled by phase separation, and that this mechanism is unchanged at least down to $\kap=10$. Experiments and simulations in Ref.~\cite{kohl2016directed} have been performed at packing fraction $\phi\simeq0.2$. We find that exactly at this packing fraction the directed percolation transition line crosses the gas-liquid binodal, leading in its vicinity to the coincidence of directed percolation and dynamic arrest. At least for the model studied here, however, no general link between the structural transition to a directed percolation network and dynamic arrest can be drawn.


\acknowledgements

Without implying their agreement with what we write, we are grateful to M. Schmiedeberg and S. Egelhaaf for critical remarks. DR acknowledges financial support by the DFG through the collaborative research center TRR 146, and CPR acknowledges the European Research Council (ERC consolidator grant NANOPRS, project number 617266).


\end{document}